\documentclass[sigconf]{acmart}
\acmConference[EASE 2026]{The 30th International Conference on Evaluation and Assessment in Software Engineering}{Tue 9 - Fri 12 June 2026}{Glasgow, United Kingdom}

\usepackage{amsmath,amsfonts}
\usepackage{algorithmic}
\usepackage{graphicx}
\usepackage{textcomp}
\usepackage{xcolor}
\usepackage{balance}
\usepackage{array}
\usepackage{adjustbox}
\usepackage{url}
\usepackage{newunicodechar}
\newunicodechar{−}{\textminus}

\newboolean{showcomments}

\setboolean{showcomments}{true}

\ifthenelse{\boolean{showcomments}}
  {
}

\definecolor{eclipseStrings}{RGB}{42,0.0,255}
\definecolor{eclipseKeywords}{RGB}{127,0,85}
\colorlet{numb}{magenta!60!black}

\def\BibTeX{{\rm B\kern-.05em{\sc i\kern-.025em b}\kern-.08em
    T\kern-.1667em\lower.7ex\hbox{E}\kern-.125emX}}
\begin{document}

\title{Bug-Report--Driven Fault Localization: Industrial Benchmarking and Lessons Learned at ABB Robotics}
%\\

\author{Pernilla Hall}
\affiliation{%
   \institution{ABB Robotics}
   %\city{L'Aquila}
   \country{Sweden}}
\email{pernilla.hall1@se.abb.com}%correggere!

\author{Anton Ununger}
\affiliation{%
   \institution{ABB Robotics}
   %\city{L'Aquila}
   \country{Sweden}}
\email{anton.ununger1@se.abb.com}%correggere!

\author{Riccardo Rubei}
\affiliation{%
   \institution{M\"{a}lardalen University}
   %\city{L'Aquila}
   \country{Sweden}}
\email{riccardo.rubei@mdu.se}

\author{Alessio Bucaioni}
\affiliation{%
   \institution{M\"{a}lardalen University}
   %\city{L'Aquila}
   \country{Sweden}}
\email{alessio.bucaioni@mdu.se}

\begin{abstract}
Software quality assurance remains a major challenge in industrial environments, where large-scale and long-lived systems inevitably accumulate defects. Identifying the location of a fault is often time-consuming and costly, particularly during maintenance phases when developers must rely primarily on textual bug reports rather than complete runtime or code-level context. In this study, we investigated whether artificial intelligence can support fault localization using only the natural-language content of bug reports. By relying exclusively on textual information, our approach requires no access to source code, execution traces, or static analysis artifacts, making it directly deployable within existing industrial maintenance workflows.

We framed fault localization as a supervised text classification problem and evaluated three traditional machine learning models (Logistic Regression, Support Vector Machine, and Random Forest) and two fine-tuned transformer-based language models (RoBERTa-Base and Distil-RoBERTa). Our evaluation used proprietary data from ABB Robotics in Västerås, Sweden, comprising approximately five years of resolved industrial bug reports, each linked to its verified code fix. This setting allowed us to assess model effectiveness under realistic industrial constraints.

Our results showed that traditional models using term frequency– inverse document frequency features consistently outperformed the fine-tuned language models on this dataset, while data augmentation substantially improved Random Forest performance. These findings challenge the assumption that transformer-based models universally outperform classical approaches in industrial contexts with domain-specific data. Overall, we demonstrated that historical bug reports can be systematically leveraged for text-based, artificial intelligence-assisted fault localization, providing a scalable, low-cost, and empirically grounded complement to traditional debugging practices in industry.

\end{abstract}

\maketitle


\section{Introduction}
\label{sec:introduction}

Modern software-intensive systems are large, interconnected, and continuously evolving. As a result, defects are inevitable and costly to diagnose and fix~\cite{SoftwareMaintenanceCosts}. A central bottleneck is fault localization, that is, identifying where a fault originates, which is time-consuming, expertise-intensive, and often performed under incomplete information~\cite{ASurveyOnSoftwareFL}. In industrial maintenance settings, developers frequently rely primarily on textual bug reports during triage, without guaranteed access to execution traces, dynamic logs, or detailed architectural context. At the same time, modern bug tracking systems accumulate large volumes of historical reports linked to verified fixes. When these reports contain sufficient diagnostic signals, the mapping between report text and fix location provides a practical foundation for Machine Learning (ML)-assisted fault localization.

In this work, we conducted a large-scale industrial study of \emph{text-only fault localization}: given a bug report (title and description), we predicted a ranked list of likely fault locations using only natural-language information. To ensure actionable recommendations, we predicted at the subfolder (component) level rather than at the individual file level. This design mirrors how developers navigate large industrial codebases and mitigates label-space sparsity. Prior research has primarily evaluated fault localization on open-source systems and often assumes access to auxiliary artifacts such as execution traces or call graphs~\cite{MultiFLML,SocialNetworkModelFL,Laprob2021106410,GHARIBI20181058,LLMsFL}. Such assumptions rarely hold in industrial maintenance contexts. Moreover, evaluations of Large Language Models (LLMs) on open-source corpora risk overlap with pretraining data, potentially inflating performance estimates. To assess feasibility under realistic industrial and confidentiality constraints, we analyzed a proprietary, large-scale ABB Robotics system in Västerås, Sweden, linking five years of resolved bug reports to their verified code changes.

We compared three traditional ML models, Logistic Regression (LR), Support Vector Machine (SVM), and Random Forest (RF), against two fine-tuned transformer-based language models, RoBERTa-Base and Distil-RoBERTa. The traditional models used term frequency–inverse document frequency (TF–IDF) representations and sentence embeddings, whereas the LLMs operated directly on tokenized reports. We applied systematic text preprocessing and class-imbalance-aware data augmentation. We evaluated all models using ranking-centric metrics (Top-$k$ Accuracy, Recall@$k$, MAP, and MRR) that reflect developer decision workflows. Our study addressed the following research questions:
\begin{enumerate}
\item[RQ1:]How do Random Forest, Logistic Regression, and Support Vector Machine perform for text-only fault localization in a proprietary industrial software system?
\item[RQ2:]How does a fine-tuned RoBERTa model (Base/Distil) compare to traditional ML models under the same industrial constraints?
\end{enumerate}

Our results showed that LR and SVM with TF–IDF features consistently outperformed more complex approaches on non-augmented data. Data augmentation substantially improved RF, and although both LLMs benefited from augmentation, they did not consistently surpass the strongest TF–IDF baselines; Distil-RoBERTa generalized more robustly than RoBERTa-Base. These findings provide concrete evidence that, in data-constrained and confidentiality-bound industrial settings, lightweight and interpretable models can match or exceed transformer-based alternatives. Despite limitations, including the single-system scope, variable report quality, and pronounced label imbalance, the evidence demonstrated that text-only ML models can reliably narrow developers’ initial search space and support cost-effective triage decisions.

This paper makes the following contributions:
\begin{itemize}
\item A large-scale industrial benchmark of text-only fault localization. We evaluated ML-assisted localization on a proprietary ABB Robotics system under realistic industrial and confidentiality constraints, mapping bug reports to verified component-level fault locations.
\item A controlled comparison of lightweight models and fine-tuned LLMs. We systematically benchmarked LR, SVM, and RF against RoBERTa-Base and Distil-RoBERTa using ranking metrics aligned with developer workflows.
\item Decision-oriented guidance for industrial adoption. We identified when simple TF–IDF-based models suffice and when the additional complexity of LLMs yields limited marginal benefit, thereby informing evidence-based technology choices.
\item Reproducible artifacts for technology transfer. We provided scripts and configuration files as a replication package (excluding proprietary data) to enable application in other industrial contexts.
\end{itemize}

Section~\ref{sec:background} positions our study within related work. Section~\ref{sec:methodology} describes the overall pipeline, and Section~\ref{sec:impl} details the experimental setup and implementation. Section~\ref{sec:results} presents the results, which we discuss in Section~\ref{sec:discussion}. Section~\ref{sec:conclusion} concludes the paper with final remarks and future research directions.

\section{Related Work}
\label{sec:background}

Fault localization has been studied through similarity-based, historical co-change, program-structure-aware, and deep learning approaches, including statistic-based methods~\cite{statistic1,statistic2,statistic3}, program state-based approaches~\cite{program1,program2,program3}, ML-based techniques~\cite{ml1,ml2,ml3}, and model-based solutions~\cite{models1,models3,models4}. Most evaluations relied on open-source datasets and assumed auxiliary artifacts such as execution traces, call graphs, stack traces, or repository navigation, which were often unavailable in industrial maintenance. In addition, LLM evaluations on public corpora risked overlap with pretraining data and could inflate performance estimates. In contrast, we evaluated text-only fault localization under realistic industrial constraints using proprietary ABB Robotics data spanning five years of resolved bug reports linked to verified fixes.

Several approaches exploited historical and structural relationships, often through graph-based representations. Chen et al.~\cite{SocialNetworkModelFL} modeled co-fixed files as an implicit social network and applied PageRank, reporting Top-10 accuracies of 32–41\%. Li et al.~\cite{Laprob2021106410} proposed LaProb, which connected reports and files via textual similarity and call graphs and propagated labels from known fixes, achieving Top-1 accuracy between 14–68\% across projects. Gharibi et al.~\cite{GHARIBI20181058} combined textual and semantic similarity, stack traces, keyword matching, and a historical multi-label Naive Bayes classifier, achieving average Top-1 $\approx$52\%, Top-10 $\approx$78\%, and MAP between 0.32 and 0.65 on three open-source systems. These results highlighted the value of combining multiple artifacts, but they depended on inputs that were not consistently accessible in industry.

Recent work explored LLM-based fault localization~\cite{xia2023conversational,wang2024software,vaithilingam2022expectation,zhang2022repairing}. Kang et al.~\cite{LLM-FL} proposed AUTOFL, where an LLM navigated a repository through function calls to retrieve code context, reporting Top-1 between $\approx$29\% and 53\%. Such approaches required interactive repository access and often depended on cloud APIs, which raised confidentiality, latency, and deployment concerns. FlexFL~\cite{10934742} introduced a two-stage framework for identifying root causes in source code and supported lightweight models within a multi-agent architecture. ChatDL~\cite{10845208} addressed change-based defect localization in flexible manufacturing software and showed that combining LLMs with information retrieval outperformed standalone GPT-based solutions. In contrast, we evaluated locally executable, text-only models without traces, call graphs, or repository interaction at inference time.

Traditional ML models have also performed well on bug report classification. Andrade et al.~\cite{ClassificationStudyML} benchmarked SVM, LR, RF, k-NN, and Naive Bayes with TF--IDF and found SVM, LR, and RF generally strongest while remaining sensitive to class imbalance. Köksal and Tekinerdogan~\cite{BilingualClassification} demonstrated industrial feasibility in a bilingual setting, achieving a weighted F1 of 72\% with a linear SVM. Other work categorized reports into technical and functional classes (e.g., Concurrency, Memory) and showed that ensembles could outperform single models despite noisy reports and class difficulty~\cite{HIRSCH2022100189}. Transformer-based classifiers such as CatIss~\cite{CatIss} further improved contextual categorization when sufficient data and metadata were available.

Unlike classification-focused studies, we addressed a more demanding \emph{multi-label ranking} problem by predicting likely components (subfolders) within a large and highly imbalanced label space. We emphasized actionable precision at top-$k$, and we compared lightweight TF--IDF baselines with fine-tuned LLMs under identical industrial constraints to provide guidance for deploying ML-assisted fault localization in real maintenance workflows.

\section{Experimental Pipeline}
\label{sec:methodology}

Our experimental pipeline comprised six stages: data collection, preprocessing, feature extraction, model selection, training, and evaluation.

\subsection{Data Collection}
We analyzed approximately five years of resolved bug reports from a large-scale ABB Robotics system (Västerås, Sweden). Each report was systematically linked to the source-code commits that fixed the issue, enabling supervised learning that maps natural-language descriptions to concrete maintenance locations in a complex, modular, and long-evolving industrial product. Using proprietary data reduced the risk of LLM pretraining overlap and provided a realistic estimate of industrial generalization. Due to confidentiality constraints, raw reports and code changes cannot be released; however, we document the full extraction and evaluation procedure to support replication on comparable industrial datasets.

\paragraph{Input Features}
For each report, we extracted two textual inputs: (i) the bug title, which typically contains domain-specific keywords indicating the affected subsystem, and (ii) the bug description, which provides contextual information such as reproduction steps, observed behavior, and occasional log fragments. We deliberately excluded non-textual metadata (e.g., severity, status, assignee) to preserve a realistic text-only deployment scenario and prevent models from exploiting process artifacts unrelated to localization.

\paragraph{Target Labels}
Target labels represented component-level fault locations at subfolder granularity. For each resolved report, we extracted modified file paths from linked commits, normalized them relative to the repository root, and mapped each file to its parent subfolder. The resulting set of subfolders formed a potentially multi-label ground truth.
We chose this abstraction for three reasons: (i) architectural coherence, as subfolders align with modules or services; (ii) developer actionability, since teams navigate code at this granularity; and (iii) learning tractability, because aggregating to subfolders substantially reduces label-space sparsity compared to file-level prediction in large industrial systems.

\subsection{Data Preprocessing}
We applied a deterministic preprocessing pipeline to standardize textual inputs. We removed reports lacking descriptions or linked fixes to ensure reliable supervision. We then applied lowercasing, punctuation removal, stop-word removal, de-camelcasing of identifiers (e.g., \texttt{RobotArmController} $\rightarrow$ “robot arm controller”), lemmatization, and tokenization. These steps reduced stylistic variability while preserving semantically relevant content for both traditional ML and transformer-based models.

\paragraph{Feature Engineering}
For traditional ML models, we used two representations: (i) TF--IDF Bag-of-Words vectors to capture lexical salience, and (ii) sentence embeddings generated with the sentence-transformers library to encode contextual similarity beyond surface forms. This enabled a controlled comparison between lexical and semantically enriched feature spaces. For transformer-based models, we tokenized reports using each model’s native tokenizer. We encoded targets as multi-hot vectors over subfolders, supporting multi-label prediction when a bug affected multiple components.

\subsection{Algorithm Selection}
Based on prior work in ML-based bug report analysis and fault localization~\cite{ClassificationStudyML,BilingualClassification,HIRSCH2022100189,CaPBugFramework,MultiLabelStudy}, and considering on-premise hardware constraints, we compared three traditional models and two transformer-based models:
\begin{itemize}
    \item RF: robust to overfitting and effective in high-dimensional spaces.
    \item LR: linear, interpretable, and a strong baseline for sparse text features.
    \item SVM: well-suited for high-dimensional and sparse text representations.
    \item RoBERTa-Base and Distil-RoBERTa: encoder-only pretrained models evaluated to balance predictive performance and computational cost.
\end{itemize}
We wrapped traditional models with \texttt{OneVsRestClassifier} to handle multi-label prediction and produce ranked outputs suitable for developer triage.

\subsection{Model Training}
We partitioned the dataset into training, validation, and test splits (70/20/10). We used the training set for model fitting, the validation set for hyperparameter tuning and early stopping (particularly during LLM fine-tuning), and the test set for a single, unbiased evaluation of generalization. We randomized splits at the bug-report level and applied multi-label stratification to preserve label distributions across partitions, preventing leakage and ensuring robust performance estimates.

\subsection{Evaluation and Comparative Analysis}
Our objective was to reduce developers’ search space by returning a ranked list of likely subfolders. We therefore evaluated models on the held-out test set using ranking-oriented, threshold-independent metrics:
\begin{itemize}
    \item Top-$k$ Accuracy (Hit@$k$) measures the proportion of reports for which at least one true label appears among the top-$k$ predictions ($k \in \{1,3,5,10\}$).
    \item Recall@$k$ measures the fraction of true labels retrieved within the top-$k$ predictions, averaged across reports.
    \item Mean Average Precision (MAP) evaluates ranking quality by averaging precision at each rank where a relevant label appears, rewarding early correct predictions.
    \item Mean Reciprocal Rank (MRR) captures how early the first relevant subfolder appears in the ranked list.
\end{itemize}

\begin{table*}[!htb]
\centering
\caption{\textbf{Synthesized example of the CSV file with extracted bugs.}}
\resizebox{\linewidth}{!}{
\begin{tabular}{|c|c|c|c|c|c|}
\hline
\textbf{Date} & \textbf{Bug ID} & \textbf{Bug Title} & \textbf{Prio.} & \textbf{Description} & \textbf{Paths} \\
\hline \hline
2021-04-02 13:57 & 123456 & Project Validation Error & 2 & Warning message when using non UTF-8 character in project name... & ../../../../File-X.cs \\
\hline
2022-01-22 14:21 & 293874 & Fatal Error When Saving & 1 & Fatal error when saving a project with non-unique name... & ../../../../File-Y.cs, ../../../../File-Z.cs \\
\hline
\end{tabular}
}
\label{tab:extracted-bugs}
\end{table*}

These metrics emphasize early precision and overall retrieval effectiveness in a multi-label setting and are standard in fault localization and bug-report analysis~\cite{Laprob2021106410,GHARIBI20181058}. By evaluating all models under identical industrial constraints, we ensured a fair and practically meaningful comparison.

\section{Experimental Setup}
\label{sec:impl}

This section describes the concrete implementation of our experimental pipeline, including environment configuration, data extraction and preparation, feature engineering, model training and tuning, and final evaluation. We ran all experiments on a Windows~11 machine using Python~3.10.9 and Visual Studio Code. We used \texttt{pandas} and \texttt{numpy} for data processing, \texttt{scikit-learn} for training and evaluating the traditional models, and \texttt{transformers} for fine-tuning and inference with RoBERTa-based models. To support reproducibility, we provide a complete, version-pinned dependency list in the replication package in Section \ref{sec:data} .

\subsection{Data Collection}
As outlined in Section~\ref{sec:methodology}, the dataset originated from a proprietary ABB Robotics software project (Västerås, Sweden). We implemented a six-step acquisition pipeline: (1) retrieve closed bug reports, (2) retain only reports linked to commits, (3) extract and aggregate modified file paths, (4) map files to subfolder-level components, (5) filter large and diffuse fixes, and (6) remove rare labels to mitigate extreme sparsity.

First, we queried the work-item database using WIQL to retrieve items of type Bug in state Closed within the project scope. This yielded 1{,}867 reports spanning approximately five years. We then queried work-item relations and retained only bugs linked to at least one Commit artifact. For the remaining reports, we extracted all modified file paths from linked commits and aggregated them per bug, producing a structured CSV file with 802 unique bug reports (see Table~\ref{tab:extracted-bugs}).

Next, we mapped each modified file path to its parent subfolder to define component-level target labels. We enumerated all unique file paths and their frequencies across commits, and we manually curated a mapping in Excel by adding a Label column corresponding to the architecturally meaningful subfolder path (see the synthesized example in Table~\ref{tab:path-mapping}). This step ensured alignment with domain-specific module boundaries rather than relying on purely syntactic path prefixes.

\begin{table}[!htb]
\centering
\caption{\textbf{Synthesized example of the CSV file with mapping between files and their corresponding subfolder.}}
\resizebox{\columnwidth}{!}{
\begin{tabular}{|c|c|c|}
\hline
\textbf{File} & \textbf{Occ.} & \textbf{Label} \\
\hline \hline
Root/Project/Folder/SubFolder-X/File-A.cs & 25 & Root/Project/Folder/SubFolder-X/ \\
\hline
Root/Project/Folder/SubFolder-Y/File-B.cs & 22 & Root/Project/Folder/SubFolder-Y/ \\
\hline
\end{tabular}
}
\label{tab:path-mapping}
\end{table}

We applied the curated mapping to append a \texttt{label\_list} column to each bug entry, capturing the unique subfolders modified for that report. We then applied two filtering steps to improve signal quality and learnability.
\begin{itemize}
    \item Large, diffuse fixes. We removed bugs linked to more than five distinct subfolders. Exploratory inspection showed that these cases typically corresponded to cross-cutting refactorings or broad maintenance changes, which dilute localization signals and distort ranking-based evaluation.
    \item Rare labels. We removed subfolder labels with fewer than 10 occurrences across the corpus and discarded bugs whose label lists became empty. This step reduced extreme class imbalance and ensured that each retained label had sufficient training support.
\end{itemize}
After filtering, the final dataset comprised 660 unique bug reports and 31 subfolder labels. The mean number of labels per bug was 1.56. The distribution of labels per report is shown in Table~\ref{tab:label-distribution}.
\begin{table}[!htb]
\centering
\caption{\textbf{Distribution of labels per bug report.}}
\begin{tabular}{|l|l|l|l|l|l|}
\hline
\textbf{Number of Labels} & 1 & 2 & 3 & 4 & 5 \\ \hline
\textbf{Number of Bugs} & 428 & 138 & 61 & 20 & 13 \\ \hline
\end{tabular}
\label{tab:label-distribution}
\end{table}

\begin{figure}[htb!]
    \centering
    \includegraphics[width=\columnwidth]{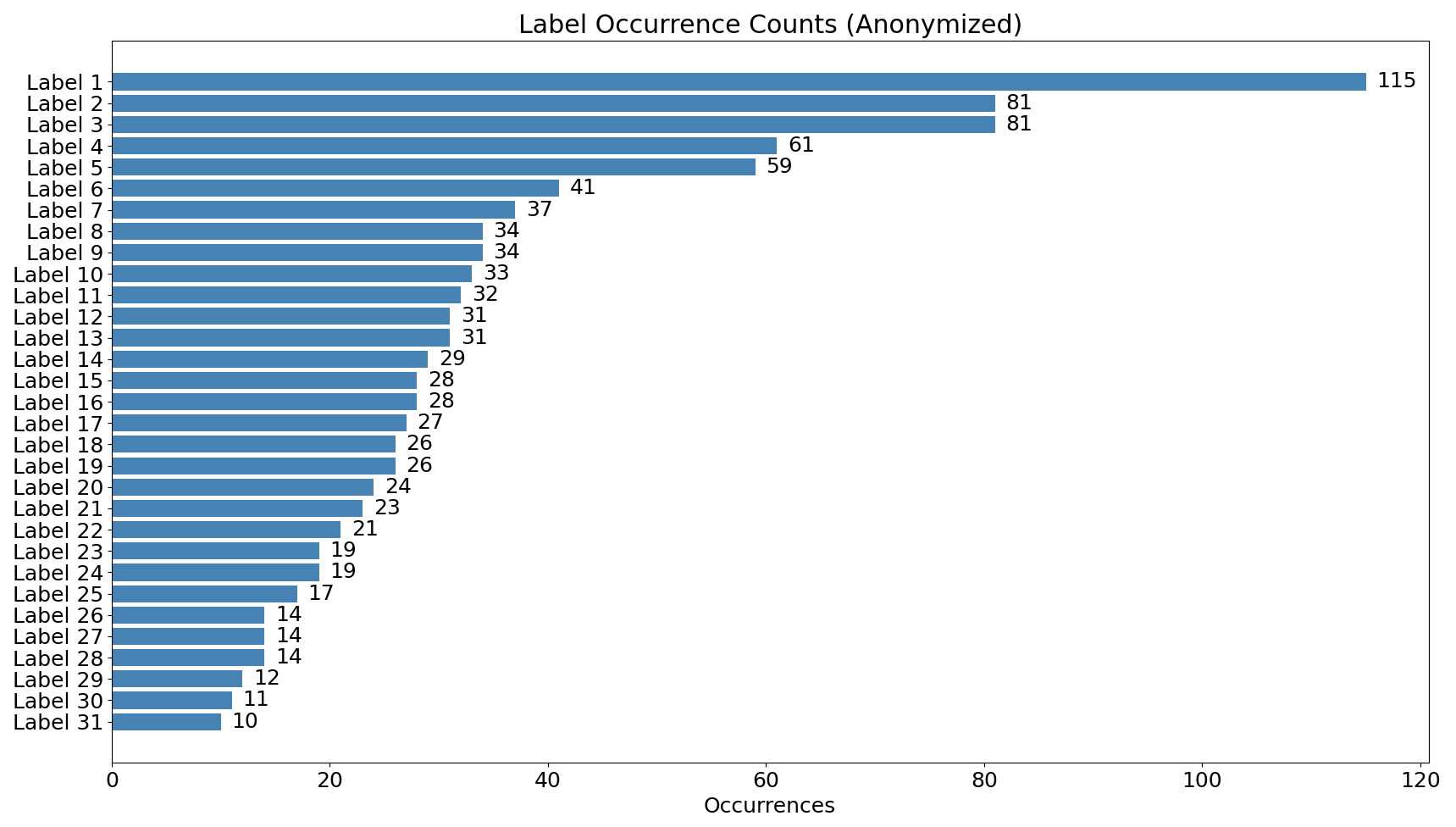}
    \caption{Visualization of label imbalance in the dataset.}
    \label{fig:class-imbalance}
\end{figure}

Prior empirical studies have shown that defect distributions often follow the Pareto principle, where a small fraction of components accounts for most faults~\cite{ParetoPrinciple}. Our dataset exhibited this pattern. As illustrated in Figure~\ref{fig:class-imbalance}, the label distribution was strongly skewed: the five most frequent subfolders occurred substantially more often than the remaining components, producing a pronounced long tail and significant class imbalance that motivated our imbalance-aware modeling strategy.

\begin{table*}[!htb]
\centering
\caption{\textbf{Optimal Random Forest hyperparameters for all datasets and features.}}
\resizebox{\linewidth}{!}{%
\begin{tabular}{|c|c|c|c|c|c|c|c|}
\hline
\textbf{Parameter} & \textbf{Values tested} & \textbf{Orig (Emb.)} & \textbf{Orig (TF)} & \textbf{RS Full} & \textbf{RS Targ.} & \textbf{SR Full} & \textbf{SR Targ.} \\
\hline \hline
TF-IDF max features & 1000, 5000, 10000, None & N/A & 1000 & 1000 & 1000 & 5000 & 1000 \\
\hline
N-gram range & (1,1), (1,2), (1,3) & N/A & (1,1) & (1,1) & (1,1) & (1,1) & (1,2) \\
\hline
Min document frequency & 1, 2, 3 & N/A & 2 & 2 & 2 & 2 & 2 \\
\hline
Number of trees & 5, 10, 20, 100 & 20 & 20 & 100 & 100 & 100 & 100 \\
\hline
Tree max depth & None, 20, 100 & None & None & None & None & None & None \\
\hline
Min samples split & 2, 5, 10 & 2 & 2 & 5 & 5 & 2 & 2 \\
\hline
Min samples leaf & 2, 5, 10, 20 & 20 & 10 & 2 & 2 & 2 & 2 \\
\hline
Class weight & balanced subsample & balanced subsample & balanced subsample & balanced subsample & balanced subsample & balanced subsample & balanced subsample \\
\hline
\end{tabular}
}
\label{tab:rf-hyperparameters}
\end{table*}

\subsection{Text Preprocessing}
We applied a deterministic natural-language preprocessing pipeline before splitting the dataset and training models. We concatenated the \emph{Bug Title} and \emph{Description} fields and normalized the resulting text using the following operations:
\begin{itemize}
    \item remove Azure DevOps template phrases (e.g., ``Detailed Description:'') using regular expressions,
    \item de-camelcase identifiers using regular expressions,
    \item case fold to lowercase,
    \item strip punctuation using regular expressions,
    \item tokenize and remove stop words using NLTK’s standard English list,
    \item lemmatize tokens using NLTK’s \texttt{WordNetLemmatizer}.
\end{itemize}
The processed dataset contained the columns \emph{processed\_text} and \emph{label\_list}, and we used it as input to dataset splitting and modeling.
%Step~6 (natural-language preprocessing) is executed with \texttt{text\_preprocess.py}.
%The resulting dataset, \texttt{bug\_dataset\_preprocessed.csv}, contains the columns \emph{processed\_text} and \emph{label\_list}, and serves as input to dataset splitting and modelling.

% \begin{figure}[hbt!]
%     \centering
%     \includegraphics[width=\columnwidth]{figs/dataset-split.png}
%     \caption{Visualization of the dataset split.}
%     \label{fig:dataset-split}
% \end{figure}
%The results of the split execution 
%The script \texttt{split\_data.py} produces:
%\begin{itemize}
%    \item \texttt{train\_df.csv}, \texttt{val\_df.csv}, \texttt{test\_df.csv}: splits containing \emph{processed\_text} and the original \emph{label\_list},
%    \item \texttt{mlb.joblib}: the fitted \texttt{MultiLabelBinarizer},
%    \item \texttt{y\_train\_bin.npy}, \texttt{y\_val\_bin.npy}, \texttt{y\_test\_bin.npy}: binarized label matrices for each split.
%\end{itemize}

%Augmentation is executed via \texttt{data\_augmentation.py}, producing four training variants in addition to the original split (Section~\ref{sec:data-split}):
%\begin{itemize}
%    \item \texttt{train\_df\_SR.csv}: full augmentation with synonym replacement,
%    \item \texttt{train\_df\_RS.csv}: full augmentation with random swaps,
%    \item \texttt{train\_df\_SR\_underrep.csv}: targeted augmentation (under-represented labels) with synonym replacement,
%    \item \texttt{train\_df\_RS\_underrep.csv}: targeted augmentation (under-represented labels) with random swaps.
%\end{itemize}

\subsection{Dataset Splitting and Label Binarization}\label{sec:data-split}
To ensure reproducibility and a like-for-like comparison across models, we performed a single randomized split (seed = 42) of the preprocessed dataset into 70\% training, 20\% validation, and 10\% test. Given the multi-label setting and class imbalance, we used \texttt{scikit-multilearn}'s \texttt{iterative\_train\_test\_split} to apply iterative stratification and approximately preserve per-label frequencies across partitions, which ensured that all labels remained represented despite the limited dataset size. After splitting, we binarized target labels (subfolder lists) into multi-hot vectors using scikit-learn’s \texttt{MultiLabelBinarizer}. We fitted the binarizer on the training labels only to prevent leakage, and then applied it to the validation and test splits.

\subsection{Data Augmentation}\label{sec:data-augmentation}
To mitigate limited data and pronounced class imbalance, we applied text-based augmentation only to the training split to increase diversity and sample counts without contaminating validation or test data. We used two techniques: (i) WordNet-based synonym replacement and (ii) random word swap, both producing semantically similar but syntactically varied reports to improve generalization.

We evaluated two augmentation strategies: (i) full augmentation of the entire training set and (ii) targeted augmentation for under-represented labels (occurrence $<25$). We used an augmentation factor of 1, generating one augmented sample per original and doubling the affected subset. This procedure yielded four additional training variants alongside the original split (Section~\ref{sec:data-split}): full synonym replacement, full random swap, targeted synonym replacement, and targeted random swap.
For each variant, we regenerated binarized labels as NumPy arrays using a MultiLabelBinarizer fitted on the corresponding training labels, resulting in five training sets in total for model training and benchmarking.

\subsection{Feature Extraction}
To prepare text into a suitable format of numerical features for traditional ML models, we test and employ two strategies: TF-IDF and sentence embeddings.

\subsection{Feature Extraction}

We transformed preprocessed text into numerical representations. For traditional ML models, we used TF--IDF vectors and sentence embeddings. For transformer-based models, we applied RoBERTa-style tokenization consistently during fine-tuning and inference.

\paragraph{TF--IDF}
We represented each report with scikit-learn’s \texttt{Tfidf\-Vectorizer}. We tuned \texttt{max\_features}, \texttt{ngram\_range}, and \texttt{min\_df} per model (Section~\ref{sec:model-training}). To prevent leakage, we fitted the vectorizer on training text only, saved it with \texttt{joblib}, and transformed the training, validation, and test splits into sparse feature matrices.

\paragraph{Sentence Embeddings}
We encoded reports with \texttt{all-mpnet-bas\-e-v2} using \texttt{sentence-transformers}. Because LR and SVM are sensitive to feature scale, we standardized embeddings with scikit-learn’s \texttt{StandardScaler} fitted on training embeddings only, saved the scaler with \texttt{joblib}, and applied it to validation and test. The scaled vectors served as inputs to the traditional ML models.

\paragraph{Tokenization}
We tokenized reports with the RoBERTa-compatible \texttt{AutoTokenizer} (Hugging Face) to produce input IDs and attention masks. We truncated and padded sequences to 256 tokens, which covered most reports in exploratory analysis while keeping compute and memory cost manageable.

\begin{table*}[!thb]
\centering
\caption{Optimal hyperparameters for RoBERTa-Base and Distill RoBERTa for all datasets.}
\resizebox{\linewidth}{!}{%
\begin{tabular}{|l|l|ll|ll|ll|ll|ll|}
\hline
\textbf{Parameter} & \multicolumn{1}{c|}{\textbf{Values tested}} & \multicolumn{2}{c|}{\textbf{Original}} & \multicolumn{2}{c|}{\textbf{RS Full}} & \multicolumn{2}{c|}{\textbf{RS Targeted}} & \multicolumn{2}{c|}{\textbf{SR Full}} & \multicolumn{2}{c|}{\textbf{SR Targeted}} \\ \hline
\begin{tabular}[c]{@{}l@{}}RoBERTa-Base / \\ Distill RoBERTa\end{tabular} & \multicolumn{1}{c|}{\textbf{Common}} & \multicolumn{1}{l|}{\textit{Base}} & \textit{Distill R.} & \multicolumn{1}{l|}{\textit{Base}} & \textit{Distill R.} & \multicolumn{1}{l|}{\textit{Base}} & \textit{Distill R.} & \multicolumn{1}{l|}{\textit{Base}} & \textit{Distill R.} & \multicolumn{1}{l|}{\textit{Base}} & \textit{Distill R.} \\ \hline
Learning rate & 1e-6 to 5e-4 & \multicolumn{1}{l|}{2.48e-05} & 1.13e−4 & \multicolumn{1}{l|}{6.95e-05} & 1.59e−4 & \multicolumn{1}{l|}{5.21e-05} & 1.73e−4 & \multicolumn{1}{l|}{3.90e-05} & 6.92e−05 & \multicolumn{1}{l|}{4.34e-05} & 1.02e−05 \\ \hline
Batch size & 4, 8, 16 & \multicolumn{1}{l|}{8} & 8 & \multicolumn{1}{l|}{16} & 8 & \multicolumn{1}{l|}{8} & 16 & \multicolumn{1}{l|}{8} & 16 & \multicolumn{1}{l|}{16} & 8 \\ \hline
Epochs & 20 & \multicolumn{1}{l|}{20} & 20 & \multicolumn{1}{l|}{20} & 20 & \multicolumn{1}{l|}{20} & 20 & \multicolumn{1}{l|}{20} & 20 & \multicolumn{1}{l|}{20} & 20 \\ \hline
Weight decay & 0.0 to 0.3 & \multicolumn{1}{l|}{0.22} & 0.21 & \multicolumn{1}{l|}{0.02} & 0.20 & \multicolumn{1}{l|}{0.24} & 0.27 & \multicolumn{1}{l|}{0.08} & 0.15 & \multicolumn{1}{l|}{0.09} & 0.09 \\ \hline
Focal Loss Alpha & 0.1 to 0.5 & \multicolumn{1}{l|}{0.38} & 0.50 & \multicolumn{1}{l|}{0.41} & 0.29 & \multicolumn{1}{l|}{0.30} & 0.30 & \multicolumn{1}{l|}{0.39} & 0.17 & \multicolumn{1}{l|}{0.36} & 0.24 \\ \hline
Focal Loss Gamma & 1 to 5 & \multicolumn{1}{l|}{4} & 2 & \multicolumn{1}{l|}{1} & 1 & \multicolumn{1}{l|}{3} & 2 & \multicolumn{1}{l|}{1} & 3 & \multicolumn{1}{l|}{2} & 5 \\ \hline
Max sequence length & 256 & \multicolumn{1}{l|}{256} & 256 & \multicolumn{1}{l|}{256} & 256 & \multicolumn{1}{l|}{256} & 256 & \multicolumn{1}{l|}{256} & 256 & \multicolumn{1}{l|}{256} & 256 \\ \hline
\begin{tabular}[c]{@{}l@{}}Gradient\\ accumulation steps\end{tabular} & 2 & \multicolumn{1}{l|}{2} & 2 & \multicolumn{1}{l|}{2} & 2 & \multicolumn{1}{l|}{2} & 2 & \multicolumn{1}{l|}{2} & 2 & \multicolumn{1}{l|}{2} & 2 \\ \hline
\end{tabular}
}
\label{tab:robertas-hyperparameters}
\end{table*}

\subsection{Model Training and Tuning}\label{sec:model-training}
\begin{figure}[!htb]
    \centering     \includegraphics[width=\columnwidth]{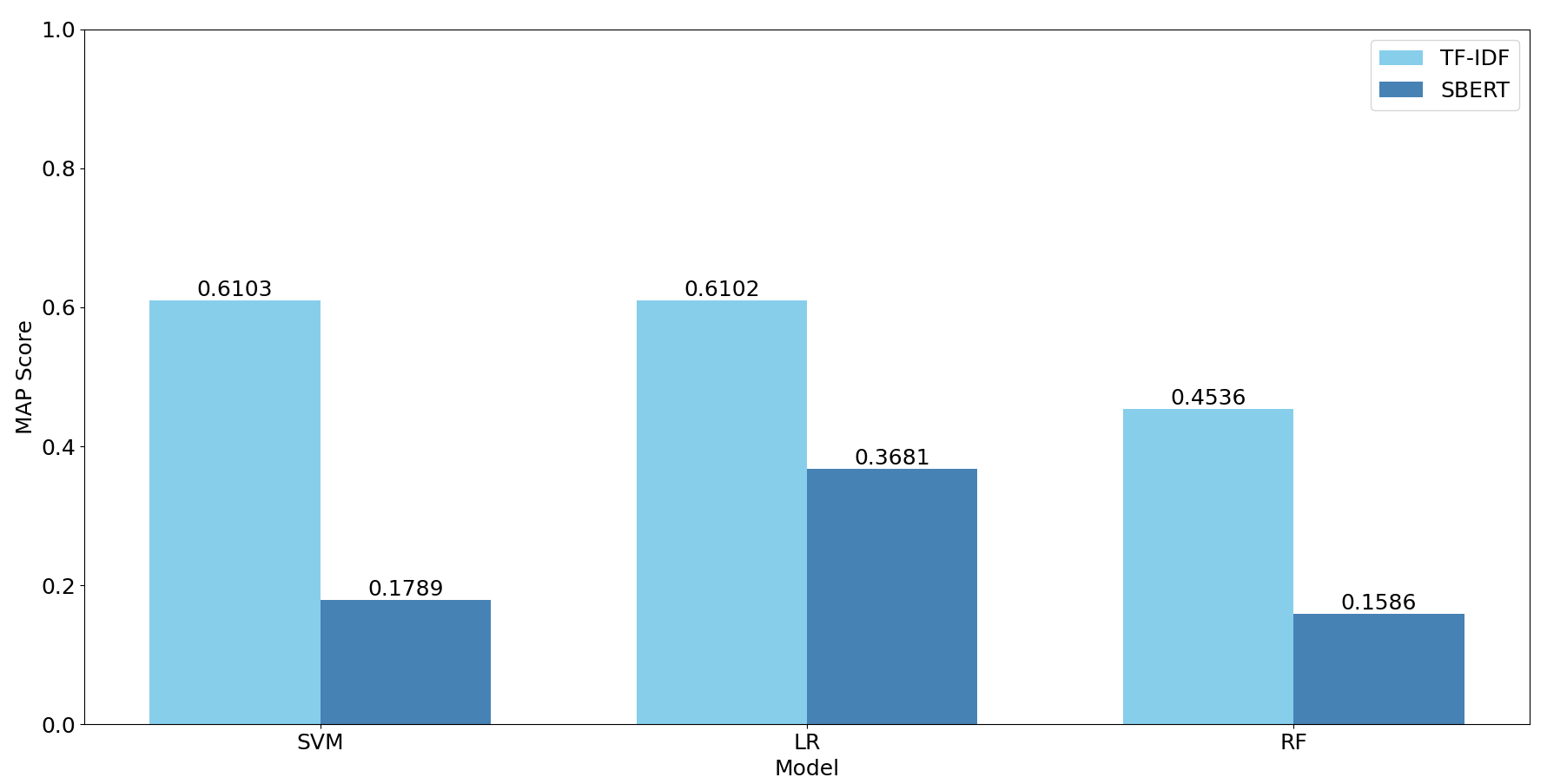}
    \caption{MAP comparison: TF--IDF vs.\ sentence embeddings for traditional models (test set).}
     \label{fig:tfidf-vs-sbert}
\end{figure}

\paragraph{Traditional ML Pipeline}
We evaluated RF, LR, and SVM under a unified pipeline with two text representations: TF--IDF vectors and sentence embeddings. For TF--IDF, we trained a scikit-learn \texttt{Pipeline} (\texttt{TfidfVectorizer} + classifier wrapped in \texttt{OneVsRest\-Classifier}) and tuned hyperparameters via grid search with 3-fold iterative stratified cross-validation on the \emph{training} split only. We selected configurations by maximizing cross-validated Mean Average Precision (MAP), refit the best model on the full training set, and saved the trained model and fitted vectorizer with \texttt{joblib}.
For embeddings, we encoded reports with \texttt{all-mpnet-base-v2} (\texttt{sentence-transformers}), standardized vectors with a \texttt{Standar\-dScaler} fitted on training embeddings only, and applied the same 3-fold iterative stratified CV and MAP-based selection. We refit the selected model on the full training set and saved the trained model and scaler.
To analyze augmentation effects, we retrained models on each augmented training variant from Section~\ref{sec:data-augmentation} while keeping validation and test splits unchanged. Because preliminary experiments showed TF--IDF outperforming embeddings, we focused augmentation analysis on the TF--IDF pipeline for controlled comparisons. Figure~\ref{fig:tfidf-vs-sbert} summarizes the TF--IDF versus embedding results.

\paragraph{Hyperparameter Selection}
We defined model-specific grids and selected hyperparameters based on cross-validated MAP. Table~\ref{tab:rf-hyperparameters} reports the explored configurations and best settings for Random Forest; we provide the full grids and results for all models in the replication package.
In Table~\ref{tab:rf-hyperparameters}, Original denotes non-augmented training data, SR Full full synonym replacement, RS Full full random swap, SR Targeted synonym replacement for under-represented labels only, and RS Targeted random swap for under-represented labels only.

\subsection{Deep Learning Pipeline}
We implemented RoBERTa-Base and Distil-RoBERTa using \texttt{Hugging\-Face Transformers} and \texttt{Datasets}, and we optimized hyperparameters with \texttt{Optuna}. For confidentiality, we pre-downloaded all models and tokenizers and executed the full pipeline locally. The pipeline comprised the following steps:
\begin{itemize}
    \item load the train/validation/test splits with multi-hot labels and wrap them as \texttt{Dataset} objects,
    \item tokenize reports with the corresponding RoBERTa tokenizer (truncation and padding to 256 tokens),
    \item fine-tune with a custom \texttt{Trainer} using Focal Loss to mitigate label imbalance,
    \item run 20 \texttt{Optuna} trials per model to sample training configurations (e.g., learning rate, batch size, weight decay, and focal-loss parameters), optimizing MAP on the validation split for comparability with traditional models,
    \item train the final model on the \emph{training} split using the best trial’s configuration, keeping the validation split fixed for selection only, and persist the model, tokenizer, and configuration for evaluation on the held-out test set.
\end{itemize}
To assess augmentation effects, we repeated the same procedure on each augmented \emph{training} variant (Section~\ref{sec:data-augmentation}) while keeping validation and test splits unchanged. Hyperparameter ranges and selected optima are reported in Table~\ref{tab:robertas-hyperparameters}.

\subsection{Final Benchmark and Evaluation Scripts}
After selecting hyperparameters on the validation split, we ran the final benchmark on the held-out test split. The evaluation script:
\begin{itemize}
    \item loaded the test set and its binarized labels,
    \item restored the best saved artifacts per model (traditional models: estimator with TF--IDF vectorizer and, where applicable, \texttt{StandardScaler}; RoBERTa models: fine-tuned model and tokenizer),
    \item generated per-label scores (probabilities/logits) and ranked subfolders accordingly for each report, and
    \item computed ranking metrics: Top-$k$ Accuracy ($k\in\{1,3,5,10\}$), Recall@$k$, Mean Reciprocal Rank (MRR), and Mean Average Precision (MAP).
\end{itemize}
We did not retrain or refit any model during this stage. All evaluations used the fixed test split under identical preprocessing. We report the results in Section~\ref{sec:results} to support a fair comparison between traditional and deep learning approaches.

\section{Results}
\label{sec:results}

We present results for the two research questions. We evaluated all configurations once on the held-out test set using the ranking metrics from Section~\ref{sec:methodology}.

\subsection{Comparison of Traditional ML Models (RQ1)}
We benchmarked LR, SVM, and RF in the text-only industrial setting using TF--IDF features. For each model, we report the strongest training regime identified during tuning (Section~\ref{sec:model-training}) and evaluated on the fixed test split: LR with Full Random Swap (RS), SVM on Original data, and RF with Full Synonym Replacement (SR). This comparison highlights how augmentation interacted with each algorithm under an identical evaluation protocol.
Overall (Table~\ref{tab:overall-best-trad}), LR+TF--IDF (RS Full) delivered the best top-ranked behavior, achieving the highest MRR and tying for Top-1 Accuracy. RF+TF--IDF (SR Full) achieved the highest MAP and Recall@1, indicating the strongest overall ranking quality with augmentation. SVM+TF--IDF (Original) led on Top-5 Accuracy and Recall@5, providing a strong baseline without augmentation.
\begin{table}[!htb]
%\begin{table}[bp]
\centering
\caption{Best traditional model configurations (test set).}
\resizebox{\columnwidth}{!}{%
\begin{tabular}{|c|c|c|c|}
\hline
Metric & LR + TF--IDF (RS Full) & SVM + TF--IDF (Original) & RF + TF--IDF (SR Full) \\
\hline \hline
Top-1 Acc.   & \textbf{0.5263} & 0.5000 & \textbf{0.5263} \\
\hline
Top-3 Acc.   & 0.7632 & 0.7632 & \textbf{0.7763} \\
\hline
Top-5 Acc.   & 0.8421 & \textbf{0.8553} & 0.7895 \\
\hline
Top-10 Acc.  & \textbf{0.9342} & 0.9211 & 0.9079 \\
\hline
Recall@1     & 0.4178 & 0.4090 & \textbf{0.4265} \\
\hline
Recall@3     & 0.6776 & 0.6919 & \textbf{0.7018} \\
\hline
Recall@5     & 0.7719 & \textbf{0.7851} & 0.7390 \\
\hline
Recall@10    & \textbf{0.8849} & 0.8805 & 0.8717 \\
\hline
MAP          & 0.6109 & 0.6103 & \textbf{0.6171} \\
\hline
MRR          & \textbf{0.6655} & 0.6526 & 0.6564 \\
\hline
\end{tabular}
}
\label{tab:overall-best-trad}
\end{table}
These results showed that TF--IDF remained highly competitive on proprietary industrial data, augmentation benefited RF the most, and LR with RS offered the strongest top-ranked recommendations. In practice, teams can prioritize LR for rapid triage, SVM for stable performance without augmentation, or RF when augmentation is feasible and overall ranking quality matters.

\subsection{Comparison of RoBERTa Models with Traditional Models (RQ2)}
We compared the best traditional TF--IDF pipelines with the best fine-tuned RoBERTa variants under identical text-only industrial constraints. As shown in Table~\ref{tab:overall-best}, neither RoBERTa-Base nor Distil-RoBERTa outperformed the strongest traditional baselines. LR with RS augmentation tied for the best Top-1 Accuracy and achieved the highest MRR, RF with SR augmentation attained the best MAP and Recall@1, and SVM without augmentation led on Top-5 Accuracy and Recall@5. Distil-RoBERTa marginally outperformed RoBERTa-Base on several metrics, but both consistently trailed the TF--IDF baselines on this limited and imbalanced dataset. Overall, the results indicated that lightweight, locally runnable TF--IDF models remained the most effective and deployable option for early-hit component ranking in industrial maintenance.
\begin{table*}[!htb]
\centering
\caption{Best configurations per model family (test set).}
\resizebox{\textwidth}{!}{%
\begin{tabular}{|c|c|c|c|c|c|}
\hline
Metric & LR+TF--IDF (RS Full) & SVM+TF--IDF (Orig.) & RF+TF--IDF (SR Full) & RoBERTa-Base (RS Targeted) & Distil-RoBERTa (RS Targeted) \\
\hline \hline
Top-1 Acc.  & \textbf{0.5263} & 0.5000 & \textbf{0.5263} & 0.4474 & 0.4605 \\
\hline
Top-3 Acc.  & 0.7632 & 0.7632 & \textbf{0.7763} & 0.7368 & 0.6974 \\
\hline
Top-5 Acc.  & 0.8421 & \textbf{0.8553} & 0.7895 & 0.8026 & 0.8026 \\
\hline
Top-10 Acc. & \textbf{0.9342} & 0.9211 & 0.9079 & 0.9079 & 0.9211 \\
\hline
Recall@1    & 0.4178 & 0.4090 & \textbf{0.4265} & 0.3662 & 0.3882 \\
\hline
Recall@3    & 0.6776 & 0.6919 & \textbf{0.7018} & 0.6458 & 0.6261 \\
\hline
Recall@5    & 0.7719 & \textbf{0.7851} & 0.7390 & 0.7292 & 0.7292 \\
\hline
Recall@10   & \textbf{0.8849} & 0.8805 & 0.8717 & 0.8465 & 0.8596 \\
\hline
MAP         & 0.6109 & 0.6103 & \textbf{0.6171} & 0.5589 & 0.5754 \\
\hline
MRR         & \textbf{0.6655} & 0.6526 & 0.6564 & 0.6028 & 0.6138 \\
\hline
\end{tabular}
}
\label{tab:overall-best}
\end{table*}

\section{Discussion}
\label{sec:discussion}
In this section, we interpret the results for the two research questions and derive implications for industrial adoption.

LR and SVM with TF--IDF features outperformed RF on the non-augmented dataset, and their near-identical results indicate that linear decision boundaries over sparse lexical features captured most of the useful signal in industrial bug reports. TF--IDF consistently surpassed sentence embeddings, showing that domain-specific terminology and recurring keywords were more predictive than generalized semantic similarity in this setting.
Augmentation produced model-dependent effects. RF benefited substantially from synonym replacement and random swaps, while LR and SVM showed marginal or slightly negative changes, likely because perturbations diluted discriminative terms. These results reinforce prior evidence that linear models with lexical features remain strong baselines for issue-text tasks~\cite{BilingualClassification,ClassificationStudyML}.

Neither RoBERTa-Base nor Distil-RoBERTa surpassed the strongest TF--IDF baselines under identical text-only constraints. Distil-RoBERTa generalized more robustly than RoBERTa-Base on the original data, consistent with expectations for smaller models under limited and imbalanced training conditions. Although augmentation improved both transformers, their absolute performance remained below LR/SVM and competitive RF.
Given the dataset size (660 reports, 31 labels) and domain-specific jargon, pretrained language models had limited opportunity to exploit broad semantic knowledge, whereas TF--IDF directly leveraged precise lexical cues aligned with component boundaries. Compared with systems such as AUTOFL~\cite{LLM-FL}, which integrate repository navigation and code context, our text-only approach prioritized deployability and confidentiality while maintaining strong ranking quality.

From an industrial perspective, deployability and cost-efficiency are central. TF--IDF pipelines (LR/SVM) and RF with augmentation provided a fully on-prem ranking aid that narrowed the search space without requiring source-code access or external APIs. On the test set, Top-1 around 0.53 (LR/RF) indicated that the first recommendation was correct in over half of cases, while Top-5 up to 0.86 and Recall@5 around 0.79 showed that shortlists typically surfaced at least one relevant component. These levels support use as an assistive triage tool rather than a fully automated locator.
A pragmatic adoption path is to start with LR or SVM using TF--IDF and introduce RF with augmentation when label imbalance becomes limiting. Integrating ranked predictions directly into bug trackers or CI dashboards maximizes impact during early investigation. Periodic retraining and monitoring of Top-$k$ and MAP help detect drift and maintain reliability. Overall, carefully tuned lightweight text-based models can deliver substantial practical value under realistic industrial constraints.

\subsection{Threats to validity}
Our findings were derived from a single proprietary ABB Robotics system, which limits external validity. Results may vary across domains, architectures, programming languages, reporting practices, and alternative label granularities. Replication on additional industrial datasets is therefore necessary to assess generalizability.
The dataset size (660 reports) and pronounced class imbalance constrained model capacity, particularly for transformer-based models. These characteristics may have reduced the potential advantage of deep architectures. In addition, confidentiality requirements mandated fully on-premise training and evaluation, which restricted exploration of larger model variants or API-based systems.
Construct validity may be influenced by our choice of subfolder-level targets as proxies for actionable components. Other organizations may prefer file-, package-, or service-level labels, which would alter task difficulty, label sparsity, and the interpretation of ranking metrics.
Internal validity risks stem from preprocessing, augmentation strategies, and hyperparameter choices. We mitigated these risks through fixed random seeds, iterative stratified splits, leakage-safe fitting procedures (e.g., fitting vectorizers, scalers, and binarizers on training data only), and consistent tuning protocols across models. Nevertheless, alternative configurations could lead to different outcomes.
For conclusion validity, we reported multiple complementary ranking metrics (Top-$k$, Recall@$k$, MAP, MRR). However, small differences between strong baselines should be interpreted cautiously given the sample size and residual noise in historical bug reports and commit links. We reduced noise by filtering incomplete entries and enforcing consistent multi-label encoding, but some inconsistencies likely remained.
Finally, the dataset is inherently confidential. All data extraction, preprocessing, model training, and evaluation were conducted locally within ABB’s secure computing environment. No proprietary data were transmitted to external servers or cloud services. Although the dataset cannot be released, we documented the full methodology, implementation steps, and evaluation procedures and provide a replication package excluding proprietary content to support methodological reproducibility on other datasets.

% \section{Related Work}
% \label{sec:relateds}
% \input{src/7_relateds}

\section{Conclusion and Future Work}
\label{sec:conclusion}

We evaluated text-only fault localization in an industrial setting using proprietary data from a large ABB Robotics system. We framed the problem as multi-label ranking over subfolders and compared three traditional models (Logistic Regression, Support Vector Machine, and Random Forest) with fine-tuned RoBERTa variants. On the held-out test set, TF--IDF-based LR and SVM consistently outperformed the transformer models, while RF became competitive when paired with simple augmentation. Sentence embeddings underperformed TF--IDF for the traditional models. The best configurations achieved Top-1 around 0.53, Top-5 up to 0.86, MAP near 0.61--0.62, and MRR around 0.66, indicating that short ranked lists can substantially narrow developers’ search space during triage. Combined with our confidentiality-friendly, fully local execution, these results support the immediate feasibility of an assistive, text-based localization capability that integrates into industrial maintenance workflows.
Our study covered a single proprietary system with 660 resolved reports and 31 subfolder labels under pronounced class imbalance. While these conditions reflect common industrial constraints, replication on additional systems is required to assess transferability across domains, architectures, and reporting practices. We provide all implementation artifacts (excluding proprietary data) as a replication package to enable reuse on comparable datasets.

We pursue three directions for future work. First, we will replicate the study across additional industrial systems and vary label granularity (file, component, and service) to understand how granularity affects accuracy and usefulness. Second, we will design hybrid, privacy-preserving models that enrich bug-report text with triage-time signals such as commit metadata, historical co-change patterns, and lightweight repository statistics, while avoiding direct ingestion of source code. Third, we will run human-in-the-loop and deployment studies to quantify effort savings (e.g., time to first correct location), calibrate confidence and abstention behavior, and integrate ranked recommendations into bug trackers and continuous integration workflows.

\section{Data Availability}
\label{sec:data}
We provide all scripts, implementation artifacts, and supporting analysis as a replication package at~\url{https://zenodo.org/records/18841230}. Due to confidentiality constraints, the underlying ABB Robotics bug reports and linked code-change data cannot be shared publicly. The package includes preprocessing, training, and evaluation scripts, configuration files, and step-by-step instructions to rerun the pipeline on comparable datasets. We do not distribute model checkpoints trained on proprietary data; instead, we provide hyperparameters and aggregated results to support reuse.

\balance

\begin{acks}This work has been partially funded by 
This work is supported by (a) the European HORIZON-KDT-JU research project MATISSE ``Model-based engineering of Digital Twins for early verification and validation of Industrial Systems", HORIZON-KDT-JU-2023-2-RIA, Proposal number:  101140216-2, KDT\-232RIA00017, (b) the Sweden's innovation agency Vinnova through the project iSecure (202301899), and (c) the Clean Energy Transition Partnership through the project ``FLEXI: Human-centered AI and digital twin powered energy system integration for flexibility markets" (101069750).
\end{acks}

\bibliographystyle{ACM-Reference-Format}
\bibliography{biblio}

@ARTICLE{ASurveyOnSoftwareFL,
  author={Wong, W. Eric and Gao, Ruizhi and Li, Yihao and Abreu, Rui and Wotawa, Franz},
  journal={IEEE Transactions on Software Engineering}, 
  title={A Survey on Software Fault Localization}, 
  year={2016},
  month={Aug.},
  volume={42},
  number={8},
  pages={707-740},
  doi={10.1109/TSE.2016.2521368}
}

@article{SoftwareMaintenanceCosts,
author = {Ogheneovo, Edward},
year = {2014},
month = {Jan.},
pages = {1-16},
title = {On the Relationship between Software Complexity and Maintenance Costs},
volume = {02},
journal = {Journal of Computer and Communications},
doi = {10.4236/jcc.2014.214001}
}

@article{MultiFLML,
author = {Gao, Meng and Li, Pengyu and Chen, Congcong and Jiang, Yunsong},
year = {2018},
month = {Nov.},
title = {Research on Software Multiple Fault Localization Method Based on Machine Learning},
volume = {232},
journal = {MATEC Web of Conferences},
doi = {10.1051/matecconf/201823201060}
}

@INPROCEEDINGS{SocialNetworkModelFL,
  author={Chen, Ing-Xiang and Yang, Cheng-Zen and Lu, Ting-Kun and Jaygarl, Hojun},
  booktitle={2008 32nd Annual IEEE International Computer Software and Applications Conference}, 
  title={Implicit Social Network Model for Predicting and Tracking the Location of Faults}, 
  year={2008},
  volume={},
  number={},
  pages={136-143},
  doi={10.1109/COMPSAC.2008.162}}

@article{LLM-FL,
author = {Kang, Sungmin and An, Gabin and Yoo, Shin},
title = {A Quantitative and Qualitative Evaluation of {LLM}-Based Explainable Fault Localization},
year = {2024},
month = {Jul.},
publisher = {Association for Computing Machinery},
address = {New York, NY, USA},
volume = {1},
issue = {FSE},
pages = {1424 - 1446},
url = {https://doi.org/10.1145/3660771},
doi = {10.1145/3660771},
journal = {Proc. ACM Softw. Eng.},
articleno = {64},
numpages = {23}
}

@article{SVM,
author = {Cortes, Corinna and Vapnik, Vladimir},
title = {Support-Vector Networks},
year = {1995},
issue_date = {Sept. 1995},
publisher = {Kluwer Academic Publishers},
address = {USA},
volume = {20},
number = {3},
issn = {0885-6125},
url = {https://doi.org/10.1023/A:1022627411411},
doi = {10.1023/A:1022627411411},
journal = {Mach. Learn.},
month = sep,
pages = {273-297},
numpages = {25}
}

@misc{ClassificationStudyML,
  author        = {Renato Andrade and César Teixeira and Nuno Laranjeiro and Marco Vieira},
  title         = "{An Empirical Study on the Classification of Bug Reports with Machine Learning}",
  howpublished  = "arXiv",
  note          = "Accessed: April 27, 2025. [Online]. Available: \url{https://arxiv.org/abs/2503.00660}"
}

@article{HIRSCH2022100189,
title = {Using textual bug reports to predict the fault category of software bugs},
journal = {Array},
volume = {15},
month = {Sept.},
year = {2022},
issn = {2590-0056},
doi = {https://doi.org/10.1016/j.array.2022.100189},
url = {https://www.sciencedirect.com/science/article/pii/S259000562200042X},
author = {Thomas Hirsch and Birgit Hofer}
}

@ARTICLE{CaPBugFramework,
  author={Ahmed, Hafiza Anisa and Bawany, Narmeen Zakaria and Shamsi, Jawwad Ahmed},
  journal={IEEE Access}, 
  title={Ca{PB}ug-{A} Framework for Automatic Bug Categorization and Prioritization Using {NLP} and Machine Learning Algorithms}, 
  year={2021},
  volume={9},
  number={},
  pages={50496-50512},
  doi={10.1109/ACCESS.2021.3069248}}

@ARTICLE{MultiLabelStudy,
  author={Zhengdong, Hu and Jantima, Polpini and Gamgarn Somprasertsri},
  journal={Journal of Information Systems Engineering and Management}, 
  title={A Comparative Study of Multilabel Classification Techniques for Analyzing Bug Report Dependencies}, 
  year={2025},
  volume={10},
  doi={10.52783/jisem.v10i26s.4253}}

@Article{BilingualClassification,
AUTHOR = {Koksal, Omer and Tekinerdogan, Bedir},
TITLE = {Automated Classification of Unstructured Bilingual Software Bug Reports: An Industrial Case Study Research},
JOURNAL = {Applied Sciences},
VOLUME = {12},
YEAR = {2022},
NUMBER = {1},
ARTICLE-NUMBER = {338},
URL = {https://www.mdpi.com/2076-3417/12/1/338},
ISSN = {2076-3417},
DOI = {10.3390/app12010338}
}

@inproceedings{LLMsFL,
author = {Yang, Aidan Z. H. and Le Goues, Claire and Martins, Ruben and Hellendoorn, Vincent},
title = {Large Language Models for Test-Free Fault Localization},
year = {2024},
isbn = {9798400702174},
publisher = {Association for Computing Machinery},
address = {New York, NY, USA},
url = {https://doi.org/10.1145/3597503.3623342},
doi = {10.1145/3597503.3623342},
booktitle = {Proceedings of the IEEE/ACM 46th International Conference on Software Engineering},
articleno = {17},
numpages = {12},
location = {Lisbon, Portugal},
series = {ICSE '24}
}

@article{GHARIBI20181058,
title = {Leveraging textual properties of bug reports to localize relevant source files},
journal = {Information Processing and Management},
volume = {54},
number = {6},
pages = {1058-1076},
year = {2018},
issn = {0306-4573},
doi = {https://doi.org/10.1016/j.ipm.2018.07.004},
url = {https://www.sciencedirect.com/science/article/pii/S0306457318301092},
author = {Reza Gharibi and Amir Hossein Rasekh and Mohammad Hadi Sadreddini and Seyed Mostafa Fakhrahmad}
}

@article{Laprob2021106410,
author = {Li, Zhengliang and Jiang, Zhiwei and Chen, Xiang and Cao, Kaibo and Gu, Qing},
year = {2020},
month = {Oct.},
title = {Laprob: A Label propagation-Based software bug localization method},
volume = {130},
journal = {Information and Software Technology},
doi = {10.1016/j.infsof.2020.106410}
}

@INPROCEEDINGS{CatIss,
  author={Izadi, Maliheh},
  booktitle={2022 IEEE/ACM 1st International Workshop on Natural Language-Based Software Engineering (NLBSE)}, 
  title={Cat{I}ss: An Intelligent Tool for Categorizing Issues Reports using Transformers}, 
  year={2022},
  volume={},
  number={},
  pages={44-47},
  doi={10.1145/3528588.3528662}}

@inproceedings{ParetoPrinciple,
author = {Walkinshaw, Neil and Minku, Leandro},
title = {Are 20\% of files responsible for 80\% of defects?},
year = {2018},
isbn = {9781450358231},
publisher = {Association for Computing Machinery},
address = {New York, NY, USA},
url = {https://doi.org/10.1145/3239235.3239244},
doi = {10.1145/3239235.3239244},
booktitle = {Proceedings of the 12th ACM/IEEE International Symposium on Empirical Software Engineering and Measurement},
articleno = {2},
numpages = {10},
location = {Oulu, Finland},
series = {ESEM '18}
}

@inproceedings{Optuna,
author = {Akiba, Takuya and Sano, Shotaro and Yanase, Toshihiko and Ohta, Takeru and Koyama, Masanori},
title = {Optuna: A Next-generation Hyperparameter Optimization Framework},
year = {2019},
isbn = {9781450362016},
publisher = {Association for Computing Machinery},
address = {New York, NY, USA},
url = {https://doi.org/10.1145/3292500.3330701},
doi = {10.1145/3292500.3330701},
booktitle = {Proceedings of the 25th ACM SIGKDD International Conference on Knowledge Discovery \& Data Mining},
pages = {2623-2631},
numpages = {9},
location = {Anchorage, AK, USA},
series = {KDD '19}
}

@ARTICLE{10934742,
  author={Xu, Chuyang and Liu, Zhongxin and Ren, Xiaoxue and Zhang, Gehao and Liang, Ming and Lo, David},
  journal={IEEE Transactions on Software Engineering}, 
  title={FlexFL: Flexible and Effective Fault Localization With Open-Source Large Language Models}, 
  year={2025},
  volume={51},
  number={5},
  pages={1455-1471},
  doi={10.1109/TSE.2025.3553363}}

@ARTICLE{10845208,
  author={Yang, Haiyang and Zhou, Yulu and Liang, Tian and Kuang, Li},
  journal={IEEE Internet of Things Journal}, 
  title={ChatDL: An LLM-Based Defect Localization Approach for Software in IIoT Flexible Manufacturing}, 
  year={2025},
  volume={12},
  number={16},
  pages={32333-32343},
  doi={10.1109/JIOT.2025.3531512}}

@article{xia2023conversational,
  title={Conversational automated program repair},
  author={Xia, Chunqiu Steven and Zhang, Lingming},
  journal={arXiv preprint arXiv:2301.13246},
  year={2023}
}

@article{wang2024software,
  title={Software testing with large language models: Survey, landscape, and vision},
  author={Wang, Junjie and Huang, Yuchao and Chen, Chunyang and Liu, Zhe and Wang, Song and Wang, Qing},
  journal={IEEE Transactions on Software Engineering},
  volume={50},
  number={4},
  pages={911--936},
  year={2024},
  publisher={IEEE}
}

@inproceedings{vaithilingam2022expectation,
  title={Expectation vs. experience: Evaluating the usability of code generation tools powered by large language models},
  author={Vaithilingam, Priyan and Zhang, Tianyi and Glassman, Elena L},
  booktitle={Chi conference on human factors in computing systems extended abstracts},
  pages={1--7},
  year={2022}
}

@article{zhang2022repairing,
  title={Repairing bugs in python assignments using large language models},
  author={Zhang, Jialu and Cambronero, Jos{\'e} and Gulwani, Sumit and Le, Vu and Piskac, Ruzica and Soares, Gustavo and Verbruggen, Gust},
  journal={arXiv preprint arXiv:2209.14876},
  year={2022}
}

@inproceedings{models1,
  author       = {Rui Abreu and
                  Arjan J. C. van Gemund},
  editor       = {Vadim Bulitko and
                  J. Christopher Beck},
  title        = {A Low-Cost Approximate Minimal Hitting Set Algorithm and its Application
                  to Model-Based Diagnosis},
  booktitle    = {Eighth Symposium on Abstraction, Reformulation, and Approximation,
                  {SARA} 2009, Lake Arrowhead, California, USA, 8-10 August 2009},
  publisher    = {{AAAI}},
  year         = {2009},
  url          = {http://www.aaai.org/ocs/index.php/SARA/SARA09/paper/view/834},
  bibsource    = {dblp computer science bibliography, https://dblp.org}
}

@inproceedings{models3,
  author       = {Vidroha Debroy and
                  W. Eric Wong},
  title        = {Insights on Fault Interference for Programs with Multiple Bugs},
  booktitle    = {{ISSRE} 2009, 20th International Symposium on Software Reliability
                  Engineering, Mysuru, Karnataka, India, 16-19 November 2009},
  pages        = {165--174},
  publisher    = {{IEEE} Computer Society},
  year         = {2009},
  url          = {https://doi.org/10.1109/ISSRE.2009.14},
  doi          = {10.1109/ISSRE.2009.14},
  bibsource    = {dblp computer science bibliography, https://dblp.org}
}

@article{models4,
  author       = {Gerhard Friedrich and
                  Markus Stumptner and
                  Franz Wotawa},
  title        = {Model-Based Diagnosis of Hardware Designs},
  journal      = {Artif. Intell.},
  volume       = {111},
  number       = {1-2},
  pages        = {3--39},
  year         = {1999},
  url          = {https://doi.org/10.1016/S0004-3702(99)00034-X},
  doi          = {10.1016/S0004-3702(99)00034-X},
  bibsource    = {dblp computer science bibliography, https://dblp.org}
}

@inproceedings{ml1,
  author       = {Luciano C. Ascari and
                  Lucilia Yoshie Araki and
                  Aurora T. R. Pozo and
                  Silvia R. Vergilio},
  title        = {Exploring machine learning techniques for fault localization},
  booktitle    = {10th Latin American Test Workshop, {LATW} 2009, Rio de Janeiro, Brazil,
                  March 2-5, 2009},
  pages        = {1--6},
  publisher    = {{IEEE}},
  year         = {2009},
  url          = {https://doi.org/10.1109/LATW.2009.4813783},
  doi          = {10.1109/LATW.2009.4813783},
  bibsource    = {dblp computer science bibliography, https://dblp.org}
}

@inproceedings{ml2,
  author       = {Lionel C. Briand and
                  Yvan Labiche and
                  Xuetao Liu},
  title        = {Using Machine Learning to Support Debugging with Tarantula},
  booktitle    = {{ISSRE} 2007, The 18th {IEEE} International Symposium on Software
                  Reliability, Trollh{\"{a}}ttan, Sweden, 5-9 November 2007},
  pages        = {137--146},
  publisher    = {{IEEE} Computer Society},
  year         = {2007},
  url          = {https://doi.org/10.1109/ISSRE.2007.31},
  doi          = {10.1109/ISSRE.2007.31},
  bibsource    = {dblp computer science bibliography, https://dblp.org}
}

@article{ml3,
  author       = {W. Eric Wong and
                  Yu Qi},
  title        = {Bp Neural Network-Based Effective Fault Localization},
  journal      = {Int. J. Softw. Eng. Knowl. Eng.},
  volume       = {19},
  number       = {4},
  pages        = {573--597},
  year         = {2009},
  url          = {https://doi.org/10.1142/S021819400900426X},
  doi          = {10.1142/S021819400900426X},
  bibsource    = {dblp computer science bibliography, https://dblp.org}
}

@article{program1,
  author       = {Andreas Zeller and
                  Ralf Hildebrandt},
  title        = {Simplifying and Isolating Failure-Inducing Input},
  journal      = {{IEEE} Trans. Software Eng.},
  volume       = {28},
  number       = {2},
  pages        = {183--200},
  year         = {2002},
  url          = {https://doi.org/10.1109/32.988498},
  doi          = {10.1109/32.988498},
  bibsource    = {dblp computer science bibliography, https://dblp.org}
}

@book{program2,
  author       = {Andreas Zeller},
  title        = {Why Programs Fail - {A} Guide to Systematic Debugging, 2nd Edition},
  publisher    = {Academic Press},
  year         = {2009},
  url          = {http://store.elsevier.com/product.jsp?isbn=9780123745156\&pagename=search},
  isbn         = {978-0-12-374515-6},
  bibsource    = {dblp computer science bibliography, https://dblp.org}
}

@inproceedings{program3,
  author       = {Holger Cleve and
                  Andreas Zeller},
  editor       = {Gruia{-}Catalin Roman and
                  William G. Griswold and
                  Bashar Nuseibeh},
  title        = {Locating causes of program failures},
  booktitle    = {27th International Conference on Software Engineering {(ICSE} 2005),
                  15-21 May 2005, St. Louis, Missouri, {USA}},
  pages        = {342--351},
  publisher    = {{ACM}},
  year         = {2005},
  url          = {https://doi.org/10.1145/1062455.1062522},
  doi          = {10.1145/1062455.1062522},
  bibsource    = {dblp computer science bibliography, https://dblp.org}
}

@inproceedings{statistic1,
  author       = {Ben Liblit and
                  Mayur Naik and
                  Alice X. Zheng and
                  Alex Aiken and
                  Michael I. Jordan},
  editor       = {Vivek Sarkar and
                  Mary W. Hall},
  title        = {Scalable statistical bug isolation},
  booktitle    = {Proceedings of the {ACM} {SIGPLAN} 2005 Conference on Programming
                  Language Design and Implementation, Chicago, IL, USA, June 12-15,
                  2005},
  pages        = {15--26},
  publisher    = {{ACM}},
  year         = {2005},
  url          = {https://doi.org/10.1145/1065010.1065014},
  doi          = {10.1145/1065010.1065014},
  bibsource    = {dblp computer science bibliography, https://dblp.org}
}

@article{statistic2,
  author       = {Chao Liu and
                  Long Fei and
                  Xifeng Yan and
                  Jiawei Han and
                  Samuel P. Midkiff},
  title        = {Statistical Debugging: {A} Hypothesis Testing-Based Approach},
  journal      = {{IEEE} Trans. Software Eng.},
  volume       = {32},
  number       = {10},
  pages        = {831--848},
  year         = {2006},
  url          = {https://doi.org/10.1109/TSE.2006.105},
  doi          = {10.1109/TSE.2006.105},
  bibsource    = {dblp computer science bibliography, https://dblp.org}
}

@article{statistic3,
  author       = {W. Eric Wong and
                  Vidroha Debroy and
                  Dianxiang Xu},
  title        = {Towards Better Fault Localization: {A} Crosstab-Based Statistical
                  Approach},
  journal      = {{IEEE} Trans. Syst. Man Cybern. Part {C}},
  volume       = {42},
  number       = {3},
  pages        = {378--396},
  year         = {2012},
  url          = {https://doi.org/10.1109/TSMCC.2011.2118751},
  doi          = {10.1109/TSMCC.2011.2118751},
  bibsource    = {dblp computer science bibliography, https://dblp.org}
}

\end{document}